\documentclass[%
 reprint,
superscriptaddress,
showpacs,
 amsmath,amssymb,
 aps,
floatfix,
]{revtex4-1}

\usepackage{graphicx}
\usepackage{dcolumn}
\usepackage{bm}

\usepackage{float}
\usepackage{amsfonts}
\usepackage{latexsym}
\usepackage{amssymb}
\usepackage{mathrsfs}
\usepackage{amsmath}
\usepackage{url} 
\usepackage{color}

\begin{document}

\preprint{APS/123-QED}
\title{Multiscale Information Storage of Linear Long-Range Correlated Stochastic Processes}


\author{Luca Faes}
 \email{luca.faes@unipa.it}
\affiliation{%
 Department of Energy, Information engineering and Mathematical models (DEIM), University of Palermo, Viale delle Scienze, Bldg. 9, 90128 Palermo, Italy
}%


\author{Margarida Almeida Pereira}
\affiliation{Faculdade de Ci\^{e}ncias, Universidade do Porto, Rua Campo Alegre, Porto, Portugal}%
\affiliation{Centro de Matem\'{a}tica da Universidade do Porto (CMUP)}%

\author{Maria Eduarda Silva}
\affiliation{%
Faculdade de Economia, Universidade do Porto, Rua Dr. Roberto Frias, Porto, Portugal}%
\affiliation{%
Centro de Investiga\c{c}\~{a}o e Desenvolvimento em Matem\'{a}tica e Aplica\c{c}\~{o}es (CIDMA)}%

\author{Riccardo Pernice}
\affiliation{%
 Department of Energy, Information engineering and Mathematical models (DEIM), University of Palermo, Viale delle Scienze, Bldg. 9, 90128 Palermo, Italy
}%

\author{Alessandro Busacca}
\affiliation{%
 Department of Energy, Information engineering and Mathematical models (DEIM), University of Palermo, Viale delle Scienze, Bldg. 9, 90128 Palermo, Italy
}%

\author{Michal Javorka}
\affiliation{%
 Department of Physiology, Comenius University in Bratislava, Jessenius Faculty of Medicine, Mala Hora 4C, 03601 Martin, Slovakia
}%
\affiliation{%
 Biomedical Center Martin, Comenius University in Bratislava, Jessenius Faculty of Medicine, Mala Hora 4C, 03601 Martin, Slovakia
}%

\author{Ana Paula Rocha}
\affiliation{Faculdade de Ci\^{e}ncias, Universidade do Porto, Rua Campo Alegre, Porto, Portugal}%
\affiliation{Centro de Matem\'{a}tica da Universidade do Porto (CMUP)}%


\date{\today}

\begin{abstract}
Information storage, reflecting the capability of a dynamical system to keep predictable information during its evolution over time, is a key element of intrinsic distributed computation, useful for the description of the dynamical complexity of several physical and biological processes. Here we introduce a parametric framework which allows to compute information storage across multiple time scales in stochastic processes displaying both short-term dynamics and long-range correlations (LRC). The framework exploits the theory of state space models to provide the multiscale representation of linear fractionally integrated autoregressive (ARFI) processes, from which information storage is computed at any given time scale relating the process variance to the prediction error variance. This enables the theoretical assessment and a computationally reliable quantification of a complexity measure which incorporates the effects of LRC together with that of short-term dynamics. The proposed measure is first assessed in simulated ARFI processes reproducing different types of autoregressive (AR) dynamics and different degrees of LRC, studying both the theoretical values and the finite sample performance. We find that LRC alter substantially the complexity of ARFI processes even at short time scales, and that reliable estimation of complexity can be achieved at longer time scales only when LRC are properly modeled. Then, we assess multiscale information storage in physiological time series measured in humans during resting state and postural stress, revealing unprecedented responses to stress of the complexity of heart period and systolic arterial pressure variability, which are related to the different role played by LRC in the two conditions.

\end{abstract}

\maketitle


\section{\label{sec:level1}Introduction}

Several physical and biological systems, such as climatic systems, econometric systems, the brain or the cardiovascular system, exhibit a rich dynamical activity that stems from the coexistence of self-sustained oscillators, interacting subsystems and feedback loops reacting to internal and external inputs \cite{donges2009complex,rosser1999complexities,chialvo2010emergent,kaplan1991aging}. This multifaceted organization results in a complex system evolution over time, which is often revealed by the time course of a systemic variable like the global temperature, the stock market, the brain wave amplitude, or the heart period. In the recent past, several techniques have been proposed which aim at quantifying the richness of a dynamic process, usually indicated as “dynamical complexity” \cite{pincus1991approximate,porta1998measuring, richman2000physiological, valente2018univariate, porta2012short, costa2002multiscale}. These methods have potentially important applications regarding both the characterization of the system state and the extraction of diagnostic parameters; for instance, a reduction of dynamical complexity may be associated with a reduced capability of subsystems to interact and, in physiological systems, has been proposed as a feature of pathologic behaviors \cite{pincus1994greater,goldberger2002physiologic}.

A common approach to assess the complexity of a time series intended as a realization of a dynamic process is that of quantifying the degree of irregularity, or unpredictability, of patterns extracted from the series. This approach has been pursued by studies proposing measures derived from linear or nonlinear prediction \cite{porta2007complexity,erla2011k,faes2015information}, or based on the concept of conditional entropy \cite{pincus1991approximate,porta1998measuring,richman2000physiological}, to quantify the dynamical complexity of a process. On the other hand a complementary measure, which has been taking place recently in the frame of information theory, is the amount of information stored in a dynamic system. The so-called information storage is defined as the information contained in the past history of a stochastic process that can be used to predict its future \cite{Lizier2012}. This measure has a straightforward information-theoretic formulation as it quantifies the information shared between the current state of a process and its past states. Moreover, besides reflecting the regularity of a dynamic process intended as a complementary measure of its complexity, this quantity is recognized as one of the three key component processes constituting every act of information processing in a network of interacting systems (i.e., information storage, transfer, and modification) \cite{wibral2014local}. As such, information storage is considered as a crucial aspect of the dynamics of several processes ranging from human brain networks \cite{kitzbichler2009broadband} to artificial networks \cite{boedecker2009initialization} and robot motion \cite{ay2008predictive}, and has been successfully proposed to describe the regularity of brain \cite{wibral2014local}, cerebrovascular \cite{faes2013investigating}, cardiorespiratory \cite{faes2015information} and cardiovascular \cite{faes2016information} dynamics.

The present work addresses the question of computing information storage across multiple temporal scales. While it is widely acknowledged that a large variety of complex systems exhibit peculiar oscillatory activities which span several different temporal scales \cite{ivanov1999multifractality,costa2002multiscale,wang2013multiscale, FaesComplexity2017}, methods to assess the information stored in a process across time scales have not been yet introduced explicitly. The most standard method to analyze the dynamics of a process at different time scales is the so-called multiscale entropy \cite{costa2002multiscale}, which is based on computing the process complexity (through the Sample Entropy measure \cite{richman2000physiological}) after eliminating the fast temporal scales through a lowpass filter and downsampling the filtered process. Although very popular, this approach suffers from some shortcomings which led to numerous refinements \cite{Valencia20092202,humeau2015multiscale}. A main issue with this measure, that is its inapplicability to short time series, has been overcome only recently with the introduction of an approach based on linear autoregressive (AR) models \cite{FaesComplexity2017}. This approach computes complexity from the parameters of a state space (SS) model which represents the rescaled version of the original AR process obtained through the filtering and downsampling steps. In this study we implement this approach for the computation of information storage and, most importantly, we extend it to account for the effect of long range correlations (LRC). LRC characterize the scaling properties displayed across time scales by a broad class of dynamic processes \cite{ivanov1999multifractality,bernaola2001scale,chen2005effect,xiong2017entropy}, and are thus a fundamental aspect of multiscale processes. Moreover, LRC are manifested also at short time scales and within the short time windows typically used for the computation of complexity measures, thus coexisting with short-term dynamics and having an impact on the assessment of their complexity \cite{xiong2017entropy}. In spite of this, methods are lacking which are able to describe quantitatively the multiscale complexity or regularity of stochastic processes in the presence of LRC. Here, we fill this gap by providing theoretical formulations and practical estimation of multiscale information storage for stochastic processes which are suitably described by fractionally integrated autoregressive (ARFI) models.

\section{Methods}
\subsection{Linear Stochastic Processes with Long Range Correlations}
We start recalling the classic parametric approach to the description of linear Gaussian stochastic processes exhibiting both short-term dynamics and long-range correlations, which is based on representing a discrete-time, zero-mean stochastic process $X_n$, $-\infty < n < \infty$, as a fractionally integrated autoregressive (ARFI) process fed by uncorrelated Gaussian innovations $E_n$. The ARFI process takes the form: 
\begin{equation} \label{eq:eq1}
		A(L)(1-L)^d X_{n} = E_{n}
\end{equation}
where \textit{L} is the back-shift operator $(L^{i}X_{n}=X_{n-i})$, $A_{L}=1-\sum\limits_{i=1}^p A_{i} L^{i}$ is an autoregressive (AR) polynomial of order \textit{p} and $(1-L)^d$ is the fractional differencing operator defined by \cite{Beran2012}:
\begin{equation} \label{eq:eq2}
		(1-L)^d = \sum\limits_{k=0}^\infty G_{k} L^{k} \quad\mathrm{,}\quad G_{k} = \frac{\Gamma(k-d)}{\Gamma(-d)\Gamma(k+1)} ,
\end{equation}
with $\Gamma(\cdot)$ denoting the gamma (generalized factorial) function. The parameter \textit{d} in (1) determines the long-term behavior of the process, while the coefficients  of the polynomial $A(L)$, i.e. 
$A_i, i=1,\ldots,p$, allow description of the short-term dynamical properties. Note that the process defined in (1) is a particular case of the broader class of ARFIMA(\textit{p,d,l}) processes, which contain also a moving average (MA) polynomial of order \textit{l} that makes  the innovation process $E_n$ non-white; here we restrict our analysis to the description of the ARFIMA(\textit{p,d},0) process, which we denote as an ARFI(\textit{p,d}) process.

The parameters of the ARFI model (1), namely the coefficients of $A(L)$ and the variance of the innovations $\Sigma _E$, are obtained from process realizations of finite length first estimating the differencing parameter $d$ by means of the Whittle semi-parametric local estimator \cite{Beran2012}, then defining the filtered data $X_{n}^{(f)}=(1-L)^{d}X_{n}$, and finally estimating the AR parameters from the filtered data $X_{n}^{(f)}$ using the ordinary least squares method to solve the AR model $A(L)X_{n}^{(f)}=E_{n}$, with model order $p$ assessed through the Bayesian information criterion \cite{faes2012measuring}.

\subsection{Linear Measures of Information Storage}
The information storage of a dynamical system that produces entropy with a non-zero rate is a quantity related to how much the system is able to share information during its evolution across time. Considering a system $\mathcal{X}$ whose activity is defined by the stochastic process $X$, let us define as $X_{n}$ the random variable describing the present state of the system, and as $X_{n}^{-} = [X_{1}X_{2}...X_{n-1}]$ the vector variable describing its past states. Then, the information stored in the system is defined as
\begin{equation} \label{eq:eq3}
		S_X=I(X_n;X_{n}^{-}) = E\left[log\frac{p(x_1,\ldots,x_n)}{p(x_1,\ldots,x_{n-1})p(x_n)}\right],
\end{equation}
where $I(\cdot;\cdot)$ denotes mutual information, $p(\cdot)$ denotes probability density function, and the expectation is taken over several realizations $(x_1,\ldots,x_{n})$ of the random variables $(X_1,\ldots,X_{n})$.

Even though information storage has been long recognized as an important aspect of the dynamics of complex systems, it has been formalized only recently as in Eq. (\ref{eq:eq3}) as the amount of information shared between the present and the past states of a dynamic process \cite{Lizier2012}. From the point of view of the dynamic update of the state of a time-evolving system, the information storage is complementary to a well-known measure of system complexity quantified in terms of entropy rate, i.e. the conditional entropy of the present state of the system given its past states, $C_{X}=H(X_n|X_{n}^{-})$ which indeed can be related to $S_{X}$ by simple information-theoretic rules: 
\begin{equation} \label{eq:eq4}
		S_X=H(X_n) - H(X_n|X_{n}^{-}) = H_X-C_X ,
\end{equation}
where $H_X = H(X_n)$ is the entropy of the present system state. Here, we exploit the relation stated in (4) and particularize it to the case of linear systems which can be fully described using an ARFI dynamic process in the form of Eq. (\ref{eq:eq1}). Specifically we note that, given the ARFI representation, the entropy of the present state of the process and the conditional entropy of the present given the past can be expressed analytically in terms of the variance of the process $X_n$, $\Sigma _X$, and the variance of the innovations $E_n$, $\Sigma _E$, as \cite{faes2015information, cover1994elements, barnett2015granger}:
\begin{subequations} \label{eq:eq5}
\begin{align}
		H(X_n) = \frac{1}{2} \ln 2\pi\mathrm{e}\Sigma{_X} \label{eq:eq5a} , \\
	  H(X_{n}|X_{n}^{-}) = \frac{1}{2} \ln 2\pi\mathrm{e}\Sigma{_{E}} \label{eq:eq5b} ,
\end{align}
\end{subequations}
from which the analytical formulation of the information storage follows immediately:
\begin{equation} \label{eq:eq6}
		S_X= \frac{1}{2} \ln \frac{\Sigma{_{X}}}{\Sigma{_{E}}} .
\end{equation}

To compute the information storage according to Eq. (\ref{eq:eq6}), we need to find an expression for the process variance $\Sigma _X$ starting from the ARFI parameters $d$ and $A(L)$ obtained as described in Sect. II.A (which allows computation also of the innovation variance $\Sigma_E$). To do this, first we approximate the ARFI process (1) with a finite order AR process by truncating the fractional integration part at a finite lag $q$:
\begin{equation} \label{eq:eq7}
		(1-L)^d \approx G(L) = \sum\limits_{k=0}^q G_{k}L^{k} ,
\end{equation}
so that the ARFI($p,d$) process can be rewritten as an AR($p+q$) process:
\begin{subequations}  \label{eq:eq8}
\begin{gather}
		B(L)X_n = E_n , \label{eq:eq8a} \\
	  B(L) = A(L)G(L) = \left(1-\sum\limits_{i=1}^p A_{i}L^{i}\right)\sum\limits_{k=0}^q G_{k}L^{k} .   \label{eq:eq8b}
\end{gather}
\end{subequations}
The coefficients of the AR polynomial $B(L)=1-\sum\limits_{k=0}^{p+q} B_{k} L^{k}$ result from the multiplication of the two polynomials in (\ref{eq:eq8b}), which in the case $q\geq p$ yields:
\begin{equation}\label{eq:eq9}\begin{split}
   B_0&=1 \quad , \\
	B_k&=\begin{cases}
   -G_k+\sum\limits_{i=1}^k G_{k-i}A_{i} ,\quad k=1,\ldots,p 	\\
	 -G_k+\sum\limits_{i=1}^p G_{k-i}A_{i} ,\quad k=p+1,\ldots,q  \\
	\sum\limits_{i=0}^{p+q-k} G_{q-i}A_{i+k-q} ,\quad k=q+1,\ldots,q+p  \\
   \end{cases}
	\end{split} .
\end{equation}
Once the ARFI process with parameters \textit{d} and \textit{p} is approximated by an AR process of order $m=p+q$, we derive the expression for the process variance using the theory of state space (SS) models \cite{barnett2015granger}. The SS formulation of the AR(\textit{m}) process of Eq. (\ref{eq:eq8}) is given by
\begin{subequations} \label{eq:eq10}
\begin{align}
		Z_{n+1} &= \mathbf{B} Z_{n} + K E_{n} \label{eq:10a} \\
		X_n &= C Z_{n} + E_{n} \label{eq:10b}
\end{align}
\end{subequations}
where $Z_{n}=[X_{n-1} \cdots X_{n-m+1} X_{n-m}]^T$ is the \textit{m}-dimensional state (unobserved) process and the vectors \textit{K} and \textit{C} and the matrix \textbf{B} are defined as:
\begin{equation}\label{eq:eq11}\begin{split}
 \mathbf{B} &= \begin{bmatrix} 
    B_{1} & \dots & B_{m-1} & B_{m} & \\
    1 & \dots & 0 & 0 & \\
		\vdots & \ddots & \vdots & \vdots & \\
    0 & \dots & 1 & 0
    \end{bmatrix} \quad , 
		K=\begin{bmatrix} 
    1 & \\
		0 & \\
		\vdots & \\
		0
    \end{bmatrix}
		\\
		C &= \begin{bmatrix} 
    B_{1} & \dots & B_{m-1} & B_{m} 
  	\end{bmatrix}
\end{split}
\end{equation}
The quantities in Eq. (\ref{eq:eq11}) are finally exploited to compute analytically the process variance $\Sigma _X$
from the solution of the following discrete-time Lyapunov equation \cite{barnett2015granger}:
\begin{equation} \label{eq:eq12}
\begin{aligned}
						\mathbf{\Omega} &= \mathbf{B}\mathbf{\Omega}\mathbf{B}^T + \Sigma{_E} K^T K \\
						\Sigma{_X} &= C\mathbf{\Omega} C^T + \Sigma{_E} ,
\end{aligned}
\end{equation}
from which the information storage is computed using Eq. (\ref{eq:eq6}).

\subsection{Multiscale Information Storage}
In this section we extend to multiple temporal scales the computation of information storage for stochastic processes which have an ARFI representation. To obtain the rescaled version of a stochastic process at the temporal scale defined by the scale factor $\tau$, the approach originally designed in \cite{costa2002multiscale} corresponds to simply take the average of the process over $\tau$ consecutive samples; this procedure has been refined later on \cite{Valencia20092202, faes2017multiscale} by recognizing that it actually entails the two subsequent steps of filtering the process with a lowpass filter with cutoff frequency $f_\tau=1/(2\tau)$, and then downsampling the filtered process using a decimation factor $\tau$. According to this refined method, we first apply a linear finite impulse response (FIR) filter to the original process $X_n$ obtaining the following filtered process:
\begin{equation} \label{eq:eq13}
\begin{aligned}
						X_{n}^{(r)} = D(L)X_n \quad ,
\end{aligned}
\end{equation}
where $r$ denotes the filter order and the filter coefficients $D(L)=\sum\limits_{k=0}^r D_{k}L^{k}$ are chosen to set up a lowpass FIR configuration with cutoff frequency $1/2\tau$. The filtering step transforms the AR(\textit{m}) process (which approximates the original ARFI(\textit{p,d}) process) into an ARMA(\textit{m,r}) process with moving average (MA) part determined by the filter coefficients:
\begin{equation} \label{eq:eq14}
\begin{aligned}
						B(L)X_{n}^{(r)} = D(L)B(L)X_{n} = D(L)E_{n}\quad .
\end{aligned}
\end{equation}
Then, we exploit the connection between ARMA processes and state space processes \cite{Aoki1991} to evidence that the ARMA process (\ref{eq:eq14}) can be expressed in SS form as:
\begin{subequations} \label{eq:eq15}
\begin{align}
		Z_{n+1}^{(r)} &= \textbf{B}^{(r)} Z_{n}^{(r)}+K^{(r)}E_{n}^{(r)} \label{eq:eq15a} \\
	  X_{n}^{(r)} &= {C}^{(r)} Z_{n}^{(r)}+E_{n}^{(r)} \label{eq:eq15b}
\end{align}
\end{subequations}
where $Z_{n}^{(r)}=[X_{n-1}^{(r)} \cdots X_{n-m}^{(r)}E_{n-1} \cdots E_{n-r}]^T$ is the $(m+r)$-dimensional state process, $E_{n}^{(r)}=D_0 E_n$ is the SS innovation process, and the vectors $K^{(r)}$ and $C^{(r)}$ and the matrix $\textbf{B}^{(r)}$ are given by: 
\begin{equation}\label{eq:eq16}\begin{split}
C^{(r)} &= \begin{bmatrix} 
    B_{1} & \cdots & B_{m}D_1 & \cdots & D_{r} 
		\end{bmatrix} \\
K^{(r)} &= \begin{bmatrix} 
    1 & \textbf{0}_{1\times(m-1)} & D_{0}^{-1} \textbf{0}_{1\times(r-1)} 
		\end{bmatrix}\\
B^{(r)} &= \begin{bmatrix} 
  C^{(r)} & \\
		 \textbf{I}_{m-1} & \textbf{0}_{(m-1)\times (r+1)} & \\
		\textbf{0}_{1\times(m+r)} & \\
		\textbf{0}_{(r-1)\times m } & \textbf{I}_{r-1} & \textbf{0}_{1\times(m+r)} & \textbf{0}_{(r-1)\times 1 } 
		\end{bmatrix}
\end{split}
\end{equation}
with $\textbf{I}_a$ and $\textbf{0}_{a\times b}$ indicating the $\textit{a}$-dimensional identity matrix and the null matrix of dimension $\textit{a} \times \textit{b}$. The parameters of the SS process are the three quantities defined in (\ref{eq:eq16}) and the variance of the innovations $\Sigma_{E^{(r)}}=D_0^{2} \Sigma_{E}$, whereby Eq. (\ref{eq:eq15}) defines an SS($\textbf{B}^{(r)},C^{(r)},K^{(r)},\Sigma_{E^{(r)}}$) process.

The second step of the rescaling procedure is to downsample the filtered process in order to complete the multiscale representation. To do this, we make use of recent theoretical findings showing that the downsampled version of an SS process has itself an SS representation \cite{solo2016state,barnett2015granger}. Here, downsampling the SS process (\ref{eq:eq15}) with a factor $\tau$ yields the process $X_{n}^{(\tau)}=X_{n\tau}^{(r)}$, which has the following SS representation:
\begin{subequations} \label{eq:eq17}
\begin{align}
		Y_{n+1} &= \textbf{B}^{(\tau)} Y_{n}+W_{n} \label{eq:eq17a} \\
	  X_{n}^{(\tau)} &= {C}^{(\tau)} Y_{n}+V_{n} \label{eq:eq17b}
\end{align}
\end{subequations}
where $\textit{V}_{n}$ and $\textit{W}_{n}$ are different white noise processes with variances $\Sigma _{W}$ and $\Sigma _{V}$ and covariance $\Sigma_{VW}$, respectively serving as innovations for the downsampled process $X_n^{(\tau)}$ and for the state process $Y_n$. Thus, the process (\ref{eq:eq17}) is an SS($\textbf{B}^{(\tau)},C^{(\tau)},\Sigma_{W},\Sigma_{V},\Sigma_{VW})$ process whose parameters can be obtained as \cite{solo2016state,barnett2015granger}:
\begin{equation} \label{eq:eq18}
\begin{aligned}
						\textbf{B}^{(\tau)} &= \left(\textbf{B}^{(r)}\right)^{\tau} \\
						C^{(\tau)} &= C^{(r)}  \\
						\Sigma_{V}&=\Sigma_{E^{(r)}} \\
						\Sigma_{VW} &= \left(\textbf{B}^{(r)}\right)^{\tau -1}{K}^{(r)}\Sigma_{E^{(r)}} \quad \\
                       \Sigma_{W}(1)&=\Sigma_{E^{(r)}} \left(K^{(r)}\right)^{T}{K}^{(r)},\quad \tau=1 
                        \\
    \Sigma_{W}(\tau)&=\textbf{B}^{(r)} \Sigma_{W}(\tau -1) \left(\textbf{B}^{(r)}\right)^{T} \\
    &+\Sigma_{E^{(r)}}\left(K^{(r)}\right)^{T}{K}^{(r)} 	,\quad \tau\geq=2 \\                   			
\end{aligned}
\end{equation}
Then, the SS($\textbf{B}^{(\tau)},C^{(\tau)},\Sigma_{W},\Sigma_{V},\Sigma_{VW}$) process can be converted in a form similar to that of Eq. (\ref{eq:eq15}) which evidences the innovations, yielding the SS ($\textbf{B}^{(\tau)},C^{(\tau)},K^{(\tau)},\Sigma_{E^{(\tau)}}$) process:
\begin{subequations} \label{eq:eq19}
\begin{align}
		Z_{n+1}^{(\tau)} &= \textbf{B}^{(\tau)} Z_{n}^{(\tau)}+K^{(\tau)}E_{n}^{(\tau)} \label{eq:eq19a} \\
	  X_{n}^{(\tau)} &= {C}^{(\tau)} Z_{n}^{(\tau)}+E_{n}^{(\tau)} \label{eq:eq19b} ,
\end{align}
\end{subequations}
which provides the SS form of the filtered and downsampled version of the original ARFI($p,d$) process. To move from (\ref{eq:eq17}) to (\ref{eq:eq19}) it is necessary to consider a discrete algebraic Ricatti equation \cite{solo2016state,barnett2015granger}:
\begin{equation} \label{eq:eq20}
\begin{aligned}
	\textbf{P}&= \textbf{B}^{(\tau)}\textbf{P}(\textbf{B}^{(\tau)})^T+\Sigma_{W}-(\textbf{B}^{(\tau)}\textbf{P}C^{(\tau)}+\Sigma_{VW})\cdot \\
    &\cdot (C^{(\tau)}\textbf{P}(C^{(\tau)})^{T}+\Sigma_{V})^{-1}(C^{(\tau)}\textbf{P}(\textbf{B}^{(\tau)})^{T}+\\&+(\Sigma_{VW})^T),
\end{aligned}
\end{equation}
which leads, after solving for \textbf{P}, to the derivation of the two remaining parameters in (\ref{eq:eq19}):
\begin{subequations} \label{eq:eq21}
\begin{align}
		\Sigma_{E^{({\tau})}} &= C^{(\tau)}\textbf{P}(C^{(\tau)})^{T}+\Sigma_{V} \label{eq:eq21a} \\
		K^{(\tau)} &= \frac{\textbf{B}^{(\tau)}\textbf{P}(C^{(\tau)})^T+\Sigma_{VW}}{\Sigma_{V}} .\label{eq:eq21b}
\end{align}
\end{subequations}
Finally, the variance of the downsampled process can be computed analytically solving a discrete-time Lyapunov equation similar to that of Eq. (\ref{eq:eq12})
\begin{subequations} \label{eq:eq22}
\begin{align}
		\mathbf{\Omega} &= \textbf{B}^{(\tau)}\mathbf{\Omega}(\textbf{B}^{(\tau)})^{T}+\Sigma_{E^{(\tau)}} (K^{(\tau)})^{T} K^{(\tau)} \label{eq:eq22a} \\
		\Sigma_{X^{(\tau)}} &= C^{(\tau)}\mathbf{\Omega}(C^{(\tau)})^{T}+\Sigma_{E^{(\tau)}} \label{eq:eq22b}
\end{align}
\end{subequations}
The derivations above lead to compute analytically the parameters of the SS process of Eq. (\ref{eq:eq19}), which constitutes a rescaled version –derived through filtering (Eq. (\ref{eq:eq15}) ) followed by downsampling (Eq. (\ref{eq:eq17}) )– of the AR approximation (Eq. (\ref{eq:eq8})) of the original ARFI process (Eq.(\ref{eq:eq1})). Among the SS parameters, the ones relevant for the computation of the information storage are the variance of the downsampled process, $\Sigma_{X^{(\tau)}}$, and the variance of the corresponding innovations, $\Sigma_{E^{(\tau)}}$. These variances can be combined in a similar way to that of Eq. (\ref{eq:eq6}) to yield the expression of the information stored in the original process $X_n$ when it is observed at scale $\tau$:
\begin{equation} \label{eq:eq23}
		S_{X}(\tau)= \frac{1}{2} \ln \frac{\Sigma{_{X^{(\tau)}}}}{\Sigma{_{E^{(\tau)}}}} .
\end{equation}

\section{Simulation study}
This section is devoted to assess the behavior of the proposed multiscale measure of information storage in stochastic processes with known dynamics. The behavior is assessed both theoretically, computing the measure for predetermined values of the parameters influencing short term dynamics and long range correlations of linear stochastic processes, and numerically, studying the performance of the adopted estimators from finite length realizations of such processes.

\subsection{Theoretical Analysis}
Here we investigate the properties of multiscale information storage by varying the parameters which determine the dynamics of ARFI processes.
These parameters are the differencing parameter $d$ and the AR coefficients composing the polynomial $A(L)$ in Eq. (\ref{eq:eq1}), which are related to long range correlations and short-term dynamics, respectively. Here, the strength of long-range correlations was varied changing the parameter $d$ in the set $\{0, 0.05, 0.4, 0.7\}$ so as to move from absent $(d=0)$ to long lasting mean reverting $(d=0.7)$ memory effects. Moreover the AR coefficients were set in order to generate stochastic oscillations with assigned frequency and spectral radius. This was achieved setting pairs of complex conjugate poles in the complex plane as the roots of the AR polynomial, where the modulus ($\rho$) or the phase ($\phi=2\pi f$, where $f$ is the frequency) of the pole was changed to reproduce varying strength and frequency of the stochastic oscillations. Two configurations were considered: (a) an AR polynomial of order $p=2$, with two poles having fixed frequency $f=0.1$ Hz and varying modulus $\rho \in \{0,0.5,0.8,0.9\}$; (b) an AR polynomial of order $p=4$ with two pairs of poles, the first with fixed modulus $\rho_1=0.8$ and frequency $f_1=0.1$ Hz, and the second with fixed modulus $\rho_2=0.8$ and varying frequency $f_2 \in \{0.15, 0.2, 0.25, 0.3\}$ Hz.

\begin{figure}[ht]
\includegraphics[width=8.5 cm]{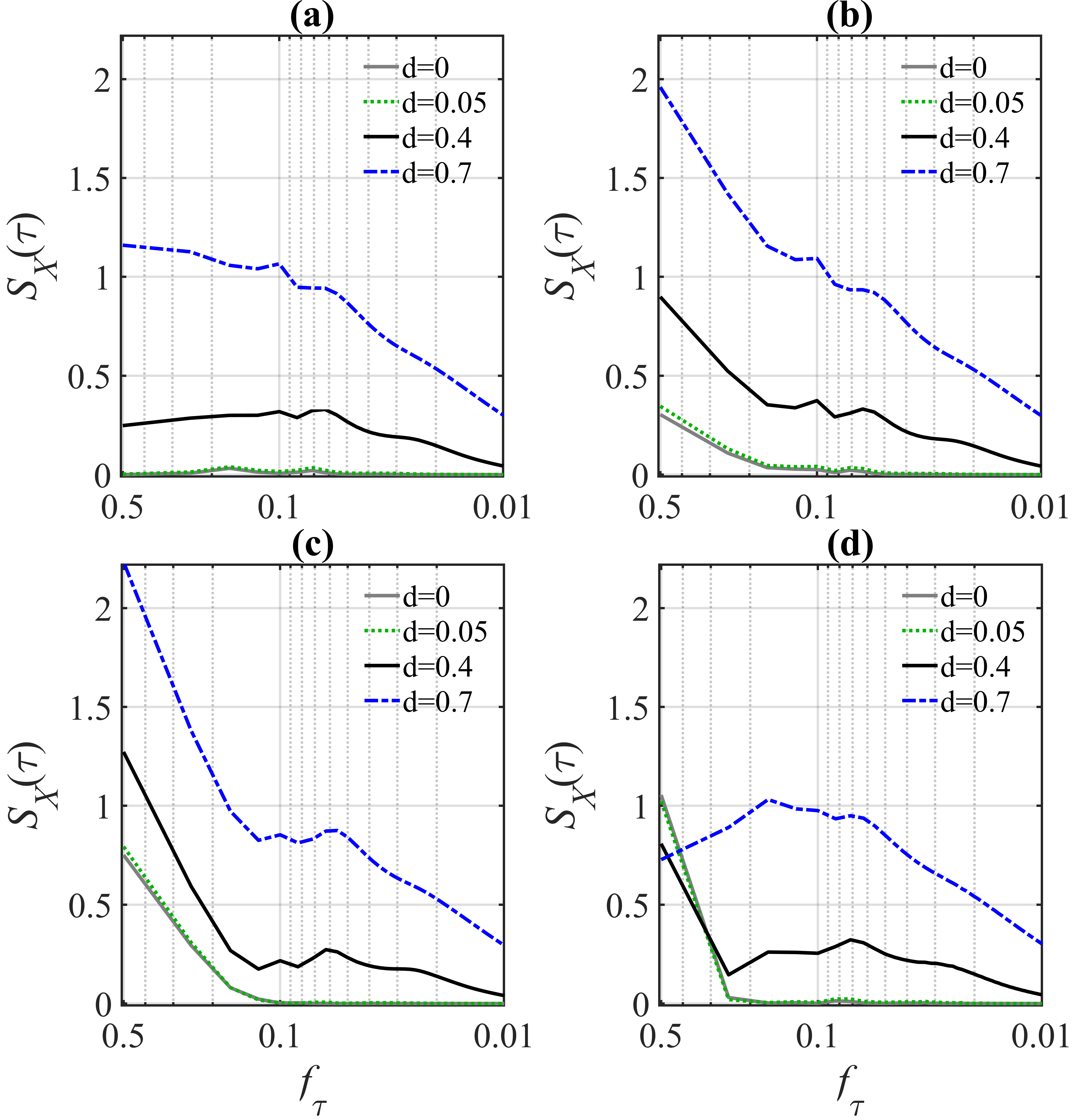}
\caption{\label{fig:simu_rho} Theoretical profiles of multiscale information storage for simulated ARFI processes with varying amplitude of stochastic oscillations. Plots depict the information storage $S_X$ computed as a function of the cutoff frequency $f_\tau$ of the lowpass filter used to change the time scale for an ARFI process characterized by two complex conjugate poles with fixed phase $\phi = 2 \pi 0.1$ and variable modulus $\rho =0$ (a), $\rho =0.5$ (b), $\rho =0.8$ (c), and $\rho =0.9$ (d), and variable differencing parameter $d=0, 0.05, 0.4, 0.7$.}
\end{figure}

\begin{figure}[ht]
\includegraphics[width=8.5 cm]{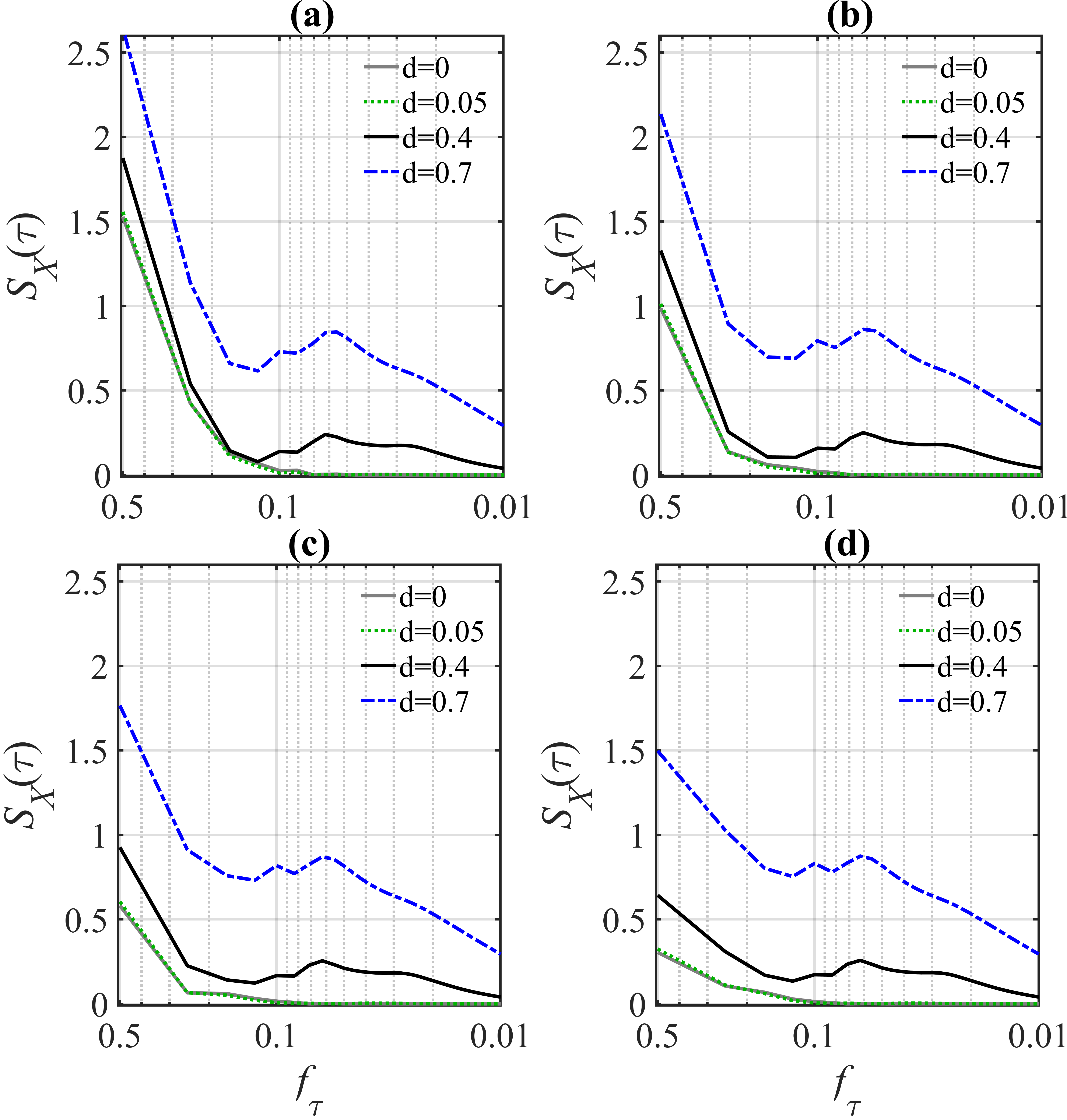}
\caption{\label{fig:simu_f} Theoretical profiles of multiscale information storage for simulated ARFI processes with varying frequency of stochastic oscillations. Plots depict the information storage $S_X$ computed as a function of the cutoff frequency $f_\tau$ of the lowpass filter used to change the time scale for an ARFI process characterized by a pair of complex conjugate poles with fixed modulus and phase ($\rho_1=0.8, \phi_1 = 2 \pi 0.1$), and another pair of complex conjugate poles with fixed modulus $\rho_2=0.8$ and variable phase $\phi_2=2 \pi f_2$, where $f_2=0.15$ (a), $f_2=0.2$ (b), $f_2=0.25$ (c), and $f_2=0.3$ Hz (d), as well as variable differencing parameter $d=0, 0.05, 0.4, 0.7$.}
\end{figure}

In each simulation, starting from the imposed theoretical values of the ARFI parameters, the analysis was performed according to the procedures described in Sect. II.B and II.C; the free parameters were set in accordance with the literature, indicating $q=50$ as an appropriate value for truncating the ARFI process to a finite order \cite{bardet2003generators}, and $r=48$ as a viable choice for the order of the lowpass filter used to implement the change of scale \cite{FaesComplexity2017}. The results, obtained for time scales increasing from 1 to 50 ($f_\tau$ decreasing from 0.5 to 0.01 Hz), are reported in Figs. \ref{fig:simu_rho},\ref{fig:simu_f}. 
In general, long range correlations tend to bring information storage into the dynamic process, to an extent proportional to the long memory of the process: indeed, for an assigned time scale $\tau$, $S_X(\tau)$ tends to increase at increasing the parameter $d$. This occurs both at the original time scale ($f_\tau=0.5$ Hz) and at longer time scales, regardless of the type of the AR process (Fig.1 and Fig. 2); the only exception is the presence of a strong stochastic oscillation ($\rho=0.9$ in Fig. \ref{fig:simu_rho}), where an increase of $d$ corresponds to a decrease of $S_X$ at the original time scale. This finding has an implication for the evaluation of entropy measures on dynamic processes in which short term dynamics coexist with long range correlations \cite{granger1980introduction,xiong2017entropy}. In these situations one should remember that, since long memory properties have an important effect on the dynamics, such properties should be accounted for to make a proper evaluation of the complexity of the observed process; if one is interested in short-term complexity only \cite{porta2012short}, long-range correlations should be removed prior to entropy analysis \cite{xiong2017entropy}.

Another general result is that the information storage tends to decrease at decreasing $f_{\tau}$, as a result of the fact that lengthening the time scale corresponds to removing regular oscillatory components and making the process more complex (i.e., less predictable); at very long time scales the process is left with no predictable dynamics and $S_X$ decays to zero. While the decrease of $S_X$ with the time scale is monotonic in the absence of long range correlations (see the curves with $d=0$ in Figs. \ref{fig:simu_rho},\ref{fig:simu_f}), the simultaneous presence of short and long memory effects may complicate the multiscale behavior of information storage. In fact, the ARFI process tends to store more information at intermediate time scales ($f_\tau \approx 0.05$ Hz) than at lower time scales when long range correlations occur simultaneously with an appreciable stochastic oscillation ($d=0.4, 0.7$ and $\rho=0.8, 0.9$, Fig. \ref{fig:simu_rho}), or with a mismatch between the frequencies of two stochastic oscillations with the same amplitude (Fig. \ref{fig:simu_f}). These patterns were not revealed by the utilization of multiscale complexity measures not accounting for long range correlations \cite{FaesComplexity2017}. Therefore, it seems that the multiscale evaluation of complexity may benefit from the use of an approach able to model dynamical effects occurring at different temporal scales such as short and long range correlations.

\begin{figure}[ht]
\includegraphics[width=8.5 cm]{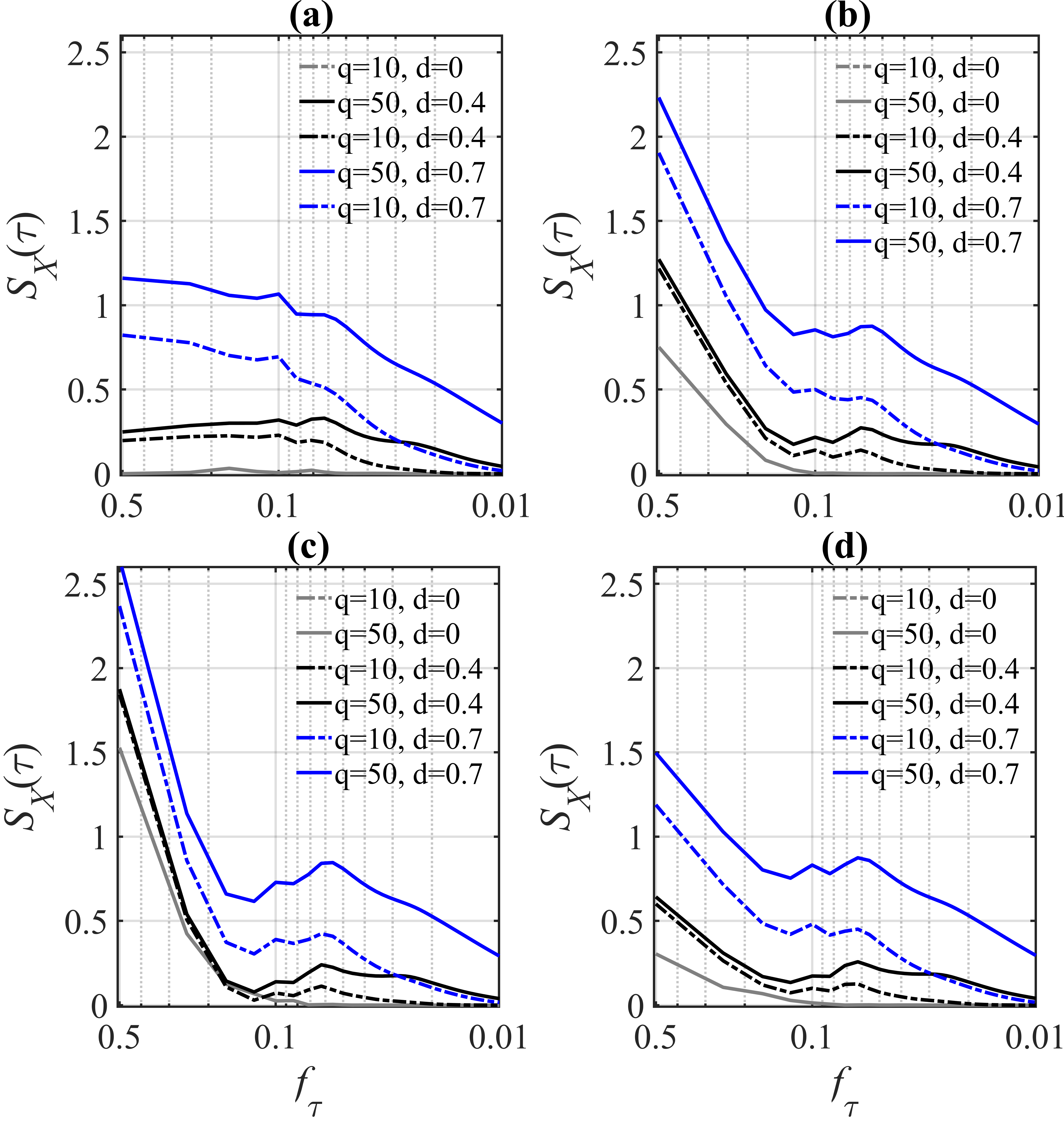}
\caption{\label{fig:simu_q}  Dependence of the theoretical profiles of multiscale information storage on the approximation of a simulated ARFI process with a finite-order AR process. Plots depict the information storage $S_X$ computed as a function of the cutoff frequency $f_\tau$ of the lowpass filter used to change the time scale for an ARFI process characterized by two complex conjugate poles with phase $\phi=2\pi 0.1$ and modulus $\rho=0$ (a) or $\rho=0.8$ (b), or by two pairs of complex conjugate poles with modulus $\rho_1=\rho_2=0.8$ and phases $\phi_1=2\pi 0.1, \phi_2=2\pi 0.15$ (c) or $\phi_1=2\pi 0.1, \phi_2=2\pi 0.3$ (d). In each panel, results are plotted for values of the differencing parameter $d=0, 0.4, 0.7$ and two values for the truncation parameter: $q=10$ (dashed lines) and $q=50$ (solid lines).}
\end{figure}

We investigated also the effects of the approximation of the ARFI process with a finite order AR process, obtained setting a fixed value for the parameter $q$ (see Eq. (\ref{eq:eq7})). Fig. \ref{fig:simu_q} reports the curves of multiscale information storage obtained in four representative AR parameter settings when assessed by the typical value $q=50$ \cite{bardet2003generators} (solid lines) and with the reduced value $q=10$ (dashed lines). Overall, we note that excessive truncation leads to an underestimation of the information storage and to smooth of non-monotonic trends of the storage with the time scale. The bias is more evident for higher values of the differencing parameter $d$ at long time scales (lower $f_\tau$). Therefore, high values of the parameter $q$ are recommended to obtain a good approximation of the long memory properties of the observed process, so to limit the negative bias of information storage and to preserve the ability to discern multiscale patterns.

\subsection{Finite Sample Performance}
Here we describe the practical estimation of multiscale information storage computed for the processes simulated as in Sect. III.A. The focus of this analysis was on assessing the computational reliability of the proposed estimator in comparison with two alternative approaches for the multiscale assessment of dynamical complexity: (i) linear multiscale analysis, based on performing pure AR identification without the modeling of long range correlations \cite{FaesComplexity2017}, and (ii) refined multiscale entropy analysis, based on computing entropy measures after the practical implementation of rescaling executed through the application of a lowpass filter followed by downsampling \cite{Valencia20092202}. The analysis was performed for a representative example of the simulation described above, where the AR polynomial was obtained from a pair of complex conjugate poles with modulus $\rho=0.8$ and frequency $f=0.1$ Hz, and for values of the fractional differencing parameter $d \in\{0, 0.4, 0.7\}$. In each analyzed case, 100 realizations of the simulation were generated deriving the polynomial $G(L)$ according to Eq. (\ref{eq:eq2}), truncating it to $q=50$ terms (Eq. (\ref{eq:eq7})), and feeding the model of Eq. (\ref{eq:eq8}) with independent samples drawn from the standard normal distribution. Then, for each realization, multiscale information storage analysis was performed for time scales $\tau=1,\ldots,50$ ($f_\tau = 0.5,\ldots,0.01$ Hz): (i) according to the procedure described in Sect. II (ARFI identification), applying a FIR lowpass filter of order $r=48$; (ii) according to linear multiscale complexity analysis, i.e. following the procedure of Sect. II but after forcing $d=0$ in Eq. (\ref{eq:eq6}) (AR identification); (iii) according to refined multiscale complexity analysis, i.e. filtering the time series with a $6^{th}$ order Butterworth lowpass filter, resampling the filtered series with a downsampling factor equal to $\tau$, and computing Sample Entropy \cite{richman2000physiological} on the downsampled time series with standard parameter setting (embedding dimension $m=2$, similarity tolerance $r=0.2 \Sigma_{E^{(\tau)}}$). To allow comparison, the complexity measures $C_X(\tau)$ derived from the complexity analyses (ii) and (iii) were converted into values of information storage exploiting the equivalence $S_X(\tau)=0.5 \ln 2\pi e - C_X(\tau)$ that holds for Gaussian processes. All estimates were compared also with the exact patterns of multiscale information storage obtained from the true values of the parameters. 

\begin{figure}[ht]
\includegraphics[width=8.5 cm]{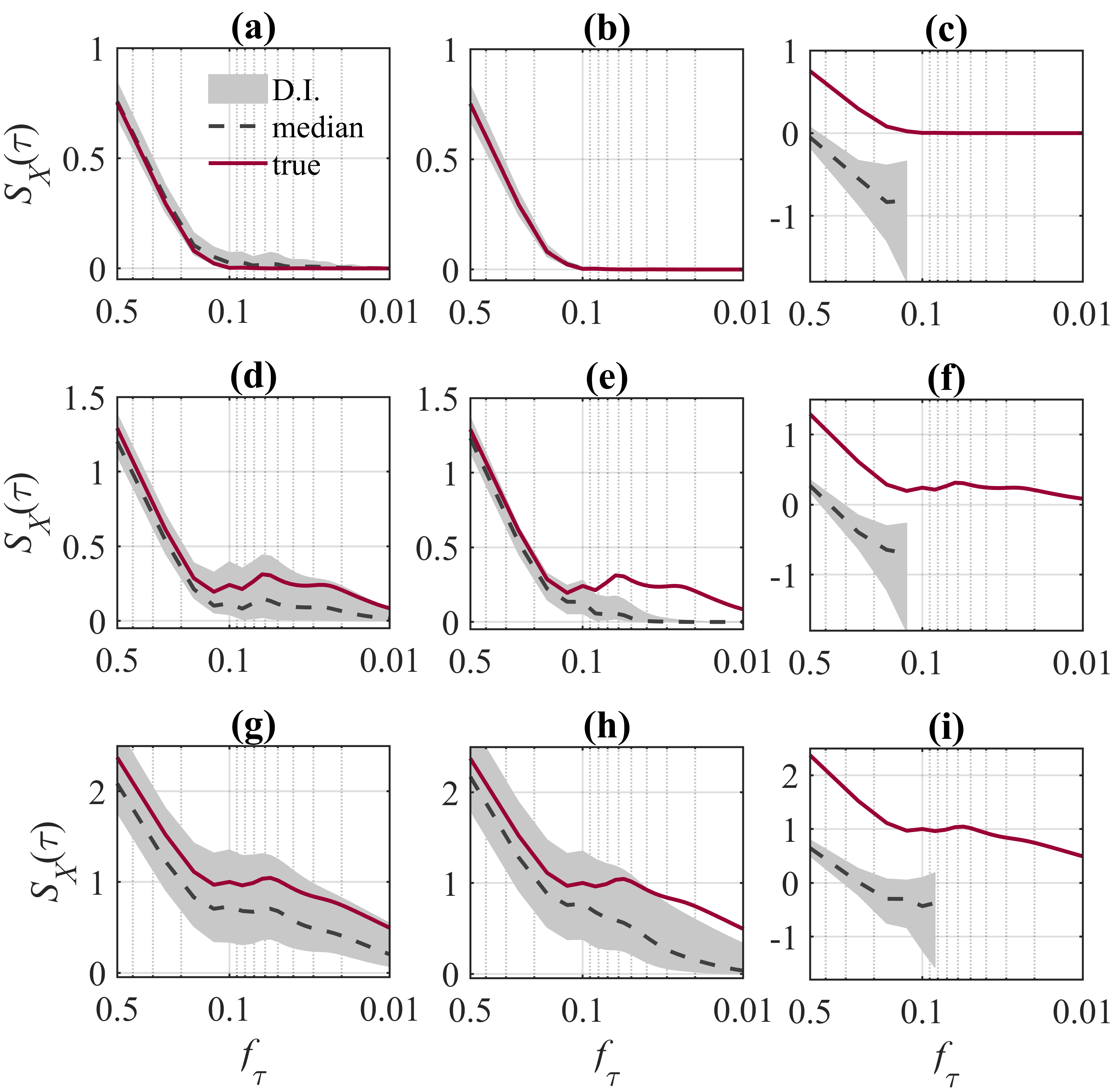}
\caption{\label{fig:simu_realization_cmp} Estimation of multiscale information storage over finite length realizations of simulated ARFI processes. Plots depict the theoretical values (red) and the distributions (median and $10^{th}-90^{th}$ percentiles (dispersion interval, D.I.) over 100 realizations) of the information storage $S_X$ computed as a function of the cutoff frequency $f_\tau$ of the lowpass filter used to change the time scale for an ARFI process characterized by two complex conjugate poles with modulus and phase $\rho=0.8, \phi=2\pi 0.1$ for values of the differencing parameters $d=0$ (a,b,c), $d=0.4$ (d,e,f) and $d=0.7$ (g,h,i), using the proposed ARFI method (a,d,g), the linear multiscale AR method \cite{FaesComplexity2017} (b,e,h), and the refined multiscale complexity approach \cite{Valencia20092202} (c,f,i).}
\end{figure}

First, we compared the distribution of $S_X(\tau)$ estimated from realizations of $N=300$ samples with its theoretical values for the three estimation approaches. The results depicted in Fig. \ref{fig:simu_realization_cmp} show that all approaches return a biased estimate of the information storage, with the bias generally increasing with the differencing parameter $d$ and with the time scale $\tau$. The bias is limited with the proposed approach based on ARFI models, as the true values of $S_X$ is contained within the dispersion interval of $10^{th}-90^{th}$ percentiles of the estimates (Fig. \ref{fig:simu_realization_cmp}a,d,g). The linear multiscale method based on pure AR identification is highly biased, in the presence of long range correlations (Fig. \ref{fig:simu_realization_cmp}e,h), at intermediate time scales ($f_\tau \le 0.1$ Hz) and becomes unreliable at longer time scales ($f_\tau \le 0.05$ Hz), returning very low values of information storage. Nevertheless, this method is highly reliable in the absence of long-range correlations (Fig. \ref{fig:simu_realization_cmp}b), performing even better than the ARFI estimator which shows a certain bias (Fig. \ref{fig:simu_realization_cmp}a); this small bias can be related to the variability in the estimation of the parameter $d$, which in turn shows non-negligible bias and variance for these estimates obtained with $N=300$ (the estimation improves for longer time series, results not shown). On the other hand, the traditional approach based on multiscale complexity analysis is highly unreliable at increasing time scales, as the estimates of $S_X$ are strongly biased, display a variance that grows dramatically with the time scale, and could not even be computed for $f_\tau \le 0.1$ Hz. These results document the necessity of the proposed approach based on ARFI models to capture the dynamical complexity of processes showing both stochastic oscillations and long memory properties.

\begin{figure}[ht]
\includegraphics[width=8.5 cm]{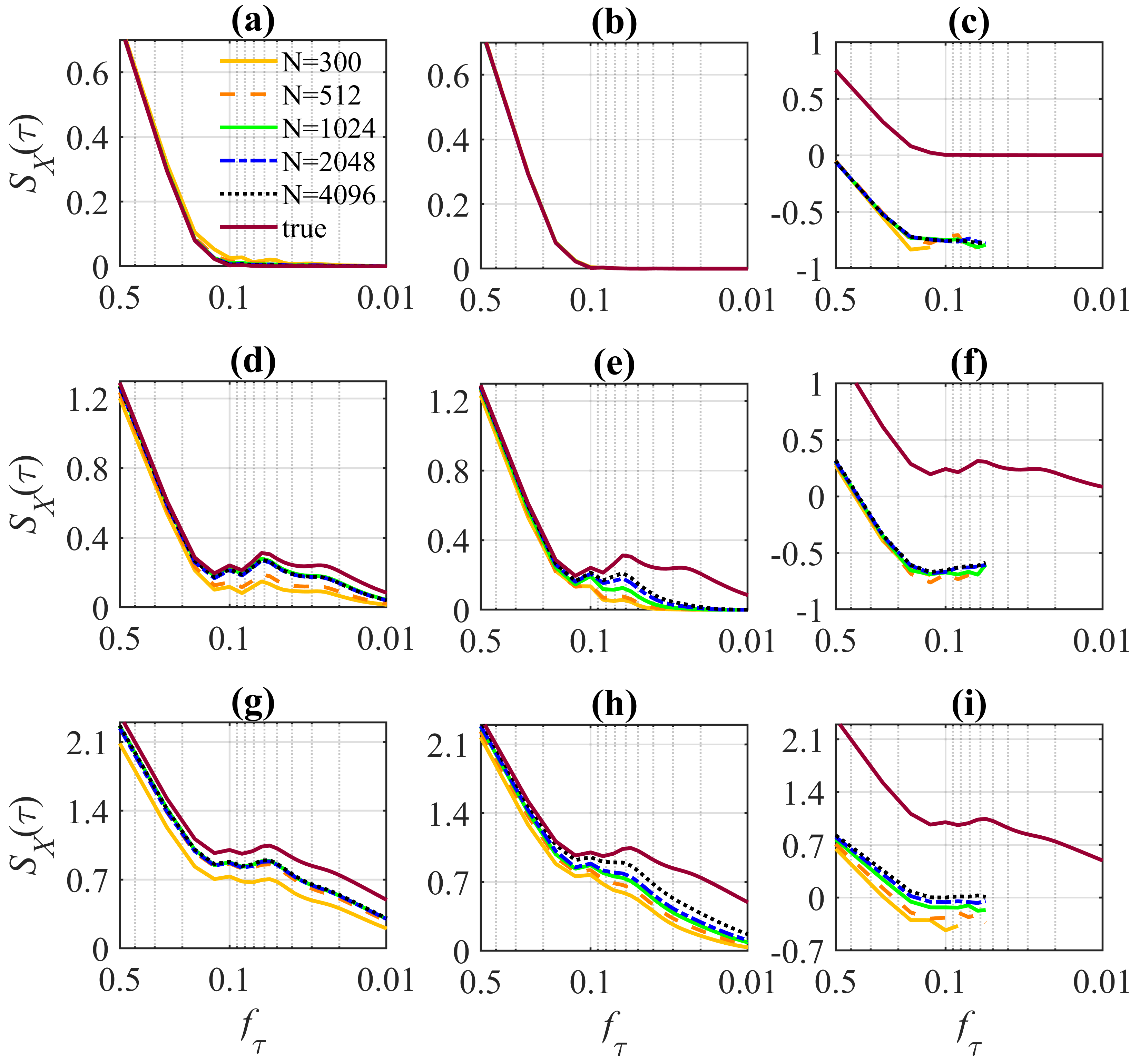}
\caption{\label{fig:simu_realization_size} Estimation of multiscale information storage as a function of the length of simulated ARFI processes. Plots depict the theoretical values (red) and the average estimated values (median over 100 realizations, other colors) of the information storage $S_X$ computed as a function of the cutoff frequency $f_\tau$ of the lowpass filter used to change the time scale for an ARFI process characterized by two complex conjugate poles with modulus and phase $\rho=0.8, \phi=2\pi 0.1$ for values of the differencing parameters $d=0$ (a,b,c), $d=0.4$ (d,e,f) and $d=0.7$ (g,h,i), using the proposed ARFI method (a,d,g), the linear multiscale AR method \cite{FaesComplexity2017} (b,e,h), and the refined multiscale complexity approach \cite{Valencia20092202} (c,f,i).}
\end{figure}

Next, we studied the dependence of the estimates of information storage on the sample size, repeating the analyses described above for time series of different length in the range $N \in \{ 300, 512, 1024, 2048, 4096 \}$. As shown in Fig. \ref{fig:simu_realization_size}, a general expected result is that the bias of $S_X$ decreases at increasing the time series length. The improvement is such that the measure based on ARFI models becomes progressively more accurate at all time scales (Fig. \ref{fig:simu_realization_size}a,d,g), while it does not help to obtain a good approximation of $S_X$ at long time scales for the measure based on AR models (Fig. \ref{fig:simu_realization_size}e,h). As to the measure based on multiscale complexity, the improvement brought by analyzing longer time series is only slight and not always clear (Fig. \ref{fig:simu_realization_size}c,f,i), confirming the unsuitability of model-free approaches to assess dynamical complexity at long time scales.

\section{Application to physiological processes}
This section reports the application of multiscale information storage in the field of cardiovascular and cardiorespiratory physiology. In this field, it is well known that the dynamics of the cardiac, vascular and respiratory systems, typically assessed from the variability series of the heart period (HP), systolic arterial pressure (SAP) and lung volume (LV), reflect the activity of physiological mechanisms operating across multiple temporal scales. In particular, the assessment of HP and SAP dynamics over temporal scales ranging from seconds to a few minutes allows the detection of short-term cardiovascular regulation, and is typically accomplished through complexity measures like approximate entropy and Sample Entropy 
\cite{pincus1991approximate, richman2000physiological}, or even using the information storage 
\cite{faes2015information,widjaja2015cardiorespiratory}. On the other hand, it is also known that cardiovascular oscillations exhibit long-range correlations properties that are manifested in scaling behavior and power law correlations which are commonly assessed using fractal techniques 
\cite{ivanov1999multifractality, bernaola2001scale}.
Given this coexistence of short-term dynamics and long-range correlations, the evaluation of the dynamical complexity of cardiovascular and respiratory processes remains a challenge that can be thoroughly faced only employing multiscale approaches. Here we investigate how the dynamical complexity of HP, SAP and LV, assessed with our measure of information storage quantified from ARFI processes, varies across multiple time scales reflecting separate but simultaneously active physiological mechanisms. Moreover we address the issue of quantifying the impact of long-range correlations, typically manifested in short cardiovascular time series in terms of slow trends superimposed to the short-term dynamics, on the values of information storage computed from short cardiovascular recordings. 

\subsection{Experimental Protocol and Measurement of Physiological Time series}
We consider the time series of HP, SAP and LV, interpreted as realizations of the stochastic processes descriptive of the cardiac, vascular and respiratory dynamics, measured in a group of 61 healthy subjects ($17.5\pm 2.4$ years old, 37 females) monitored in the resting supine position (SU) and in the upright position (UP) reached through passive head-up tilt \cite{javorka2017causal}. The acquired signals were the surface electrocardiogram (ECG), the finger arterial blood pressure recorded noninvasively by the photoplethysmographic method, and the respiration signal recorded through respiratory inductive plethysmography. For each subject and experimental condition, the values of HP, SAP and LV were measured on a beat-to-beat basis respectively as the sequences of the temporal distances between consecutive R peaks of the ECG, the maximum values of the arterial pressure waveform taken within the consecutively detected heart periods, and the values of the respiratory signal sampled at the onset of the consecutively detected heart periods. A detailed description of experimental protocol and signal measurement is reported in Ref. \cite{javorka2017causal}.

The analysis was performed on segments of $N=300$ consecutive points, free of artifacts and deemed as weak-sense stationary through visual inspection, extracted from the time series for each subject and condition. Three different approaches were followed to compute multiscale information storage: (i) the “eAR” approach, based on pure AR model identification, i.e. performing the whole procedure described in Sect. II after forcing \textit{d}=0 in Eq. (\ref{eq:eq6}); (ii) the “eARd” approach, based on pure AR identification as in (i), but applied to the filtered data $X_{n}^{(f)}=(1-L)^{d}X_n$, after estimating the parameter \textit{d} from the original time series;
(iii) the “eARFI” approach, based on complete ARFI model identification, i.e., following the whole procedure described in Sect. II with \textit{d} estimated from the original time series and considered in the computations. Pursuing these approaches we compare, respectively, (i) the traditional complexity analysis where long-range correlations are neither removed nor modeled, (ii) the analysis performed only on the short-term dynamics after removing long-range correlations, and (iii) the complexity analysis performed by modeling the long range correlations and considering them together with the short-term dynamics. Such a comparison is meant to infer the role of long-range correlations vs. that of short-term dynamics in contributing to the information storage and to its variation between conditions.

The ARFI model fitting each individual time series was identified first estimating the fractional differencing parameter \textit{d} using the Whittle estimator, then filtering the time series with the fractional integration polynomial truncated at a lag $\textit{q}=50$, and finally estimating the parameters of the polynomial relevant to the short-term dynamics through least squares AR identification. The AR model order $p$ was selected as the minimum of the BIC figure of merit \cite{stoica2004model} in the range 2-16.  Then, multiscale information storage was computed implementing a FIR lowpass filter of order $\textit{r}=48$, for time scales $\tau$ in the range (1,\dots,400), which corresponds to lowpass cutoff frequencies $f_{\tau}=(0.5,...,0.00125)$ Hz.

Here the effects of SU and UP conditions on the information storage profiles are assessed at any assigned time scale by means of paired comparisons.

\subsection{Results and Discussion}
\begin{figure}[ht]
\includegraphics[width=8.5 cm]{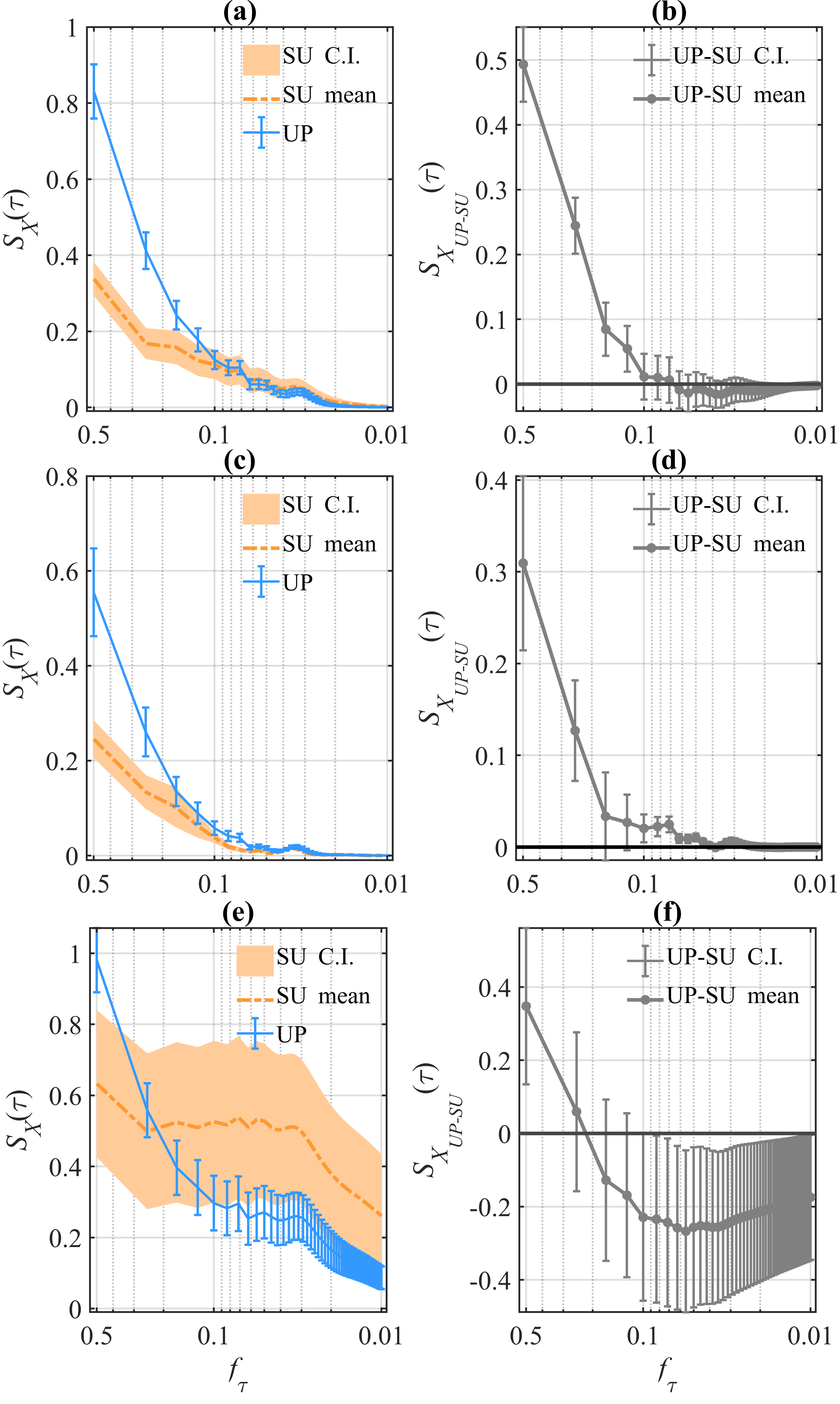}
\caption{\label{fig:HPresults} Multiscale Information Storage for the heart period time series. Plots report the confidence interval (C.I.) for the mean of the index of information storage computed across subjects using the eAR approach (a), the eARd method (c) and the eARFI method (e) as a function of the cutoff frequency of the rescaling filter in the supine (SU) and upright (UP) body positions. For each estimation method, the paired C.I. of UP$-$SU are also plotted in panels (b,d,f).}
\end{figure}

\begin{figure}[ht]
\includegraphics[width=8.5 cm]{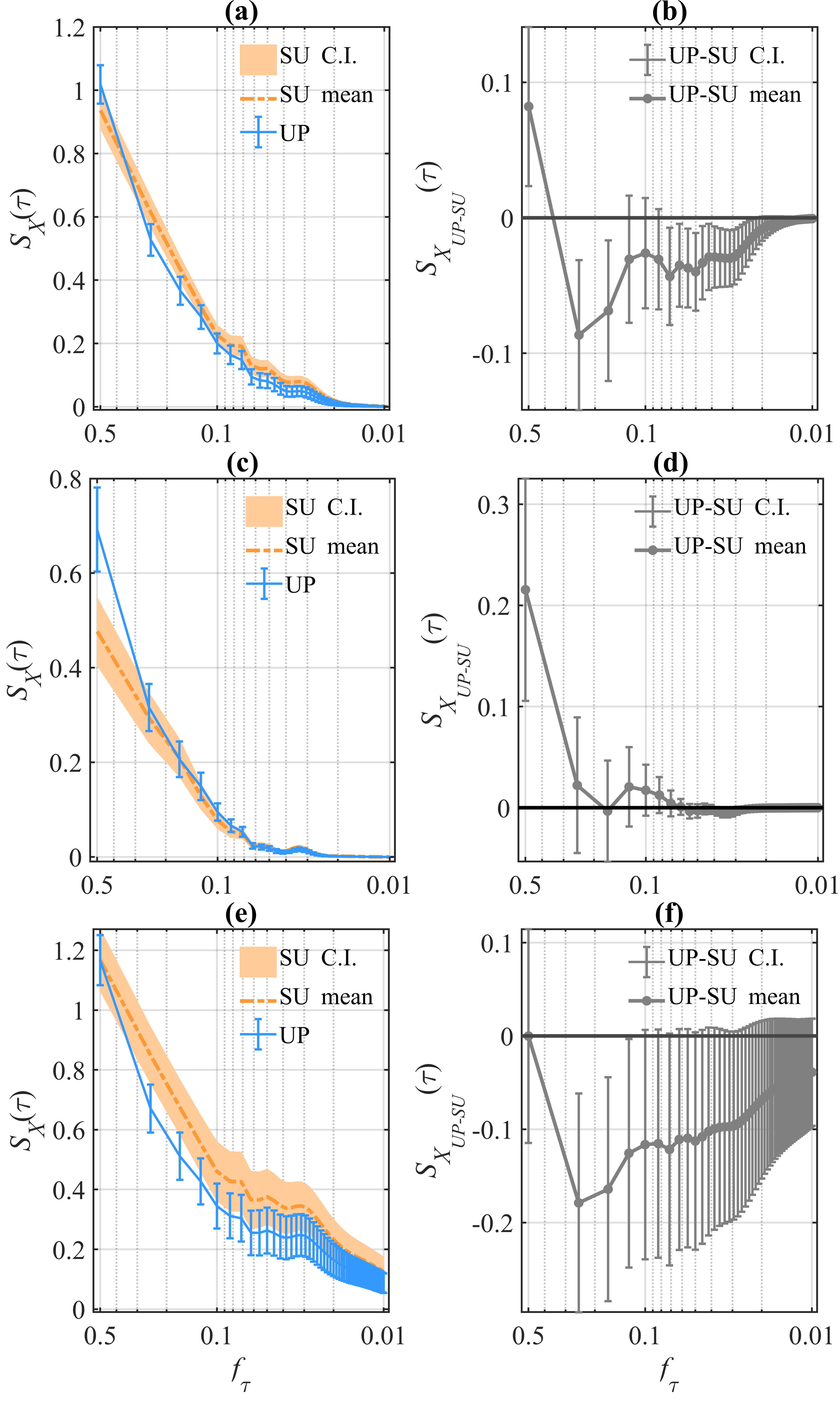}
\caption{\label{fig:SAPresults} Multiscale Information Storage for the systolic pressure time series. Plots report the confidence interval (CI) for the mean of the index of information storage computed across subjects using the eAR approach (a), the eARd method (c) and the eARFI method (e) as a function of the cutoff frequency of the rescaling filter in the supine (SU) and upright (UP) body positions. For each estimation method, the paired CI of UP$-$SU are also plotted in panels (b,d,f).}
\end{figure}

\begin{figure}[ht]
\includegraphics[width=8.5 cm]{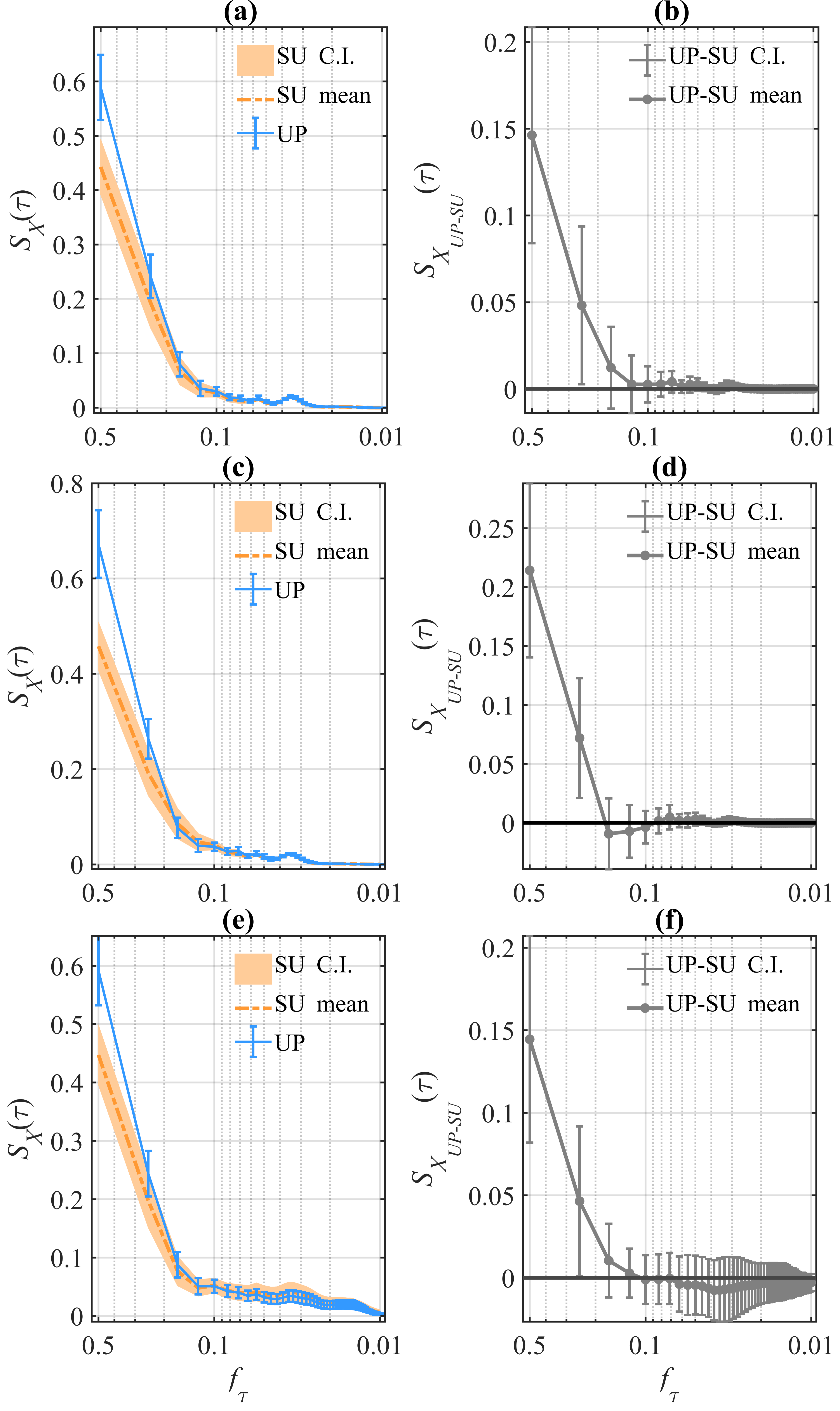}

\caption{\label{fig:RESPresults}{Multiscale Information Storage for the respiratory time series. Plots report the confidence interval (CI) for the mean of the index of information storage computed across subjects using the eAR approach (a), the eARd method (c) and the eARFI method (e) as a function of the cutoff frequency of the rescaling filter in the supine (SU) and upright (UP) body positions. For each estimation method, the paired CI of UP$-$SU are also plotted in panels (b,d,f).}}

\end{figure}

The results of the multiscale computation of information storage for the HP, SAP and LV time series are depicted respectively in Figs. \ref{fig:HPresults},  \ref{fig:SAPresults}, and \ref{fig:RESPresults}, reporting the distribution across subjects of the index $S_X$ (left column) computed following the eAR (first row), eARd (second row) and eARFI (third row) estimation approaches and evaluated as a function of the time scale in the two analyzed physiological states (SU and UP). In each figure, results of the statistical analysis are also visualized (right column) reporting the mean and $95\%$ confidence intervals of the paired difference between the values of $S_X$ computed in the UP and SU conditions; a statistically significant variation from SU to UP is detectable at a given time scale if the confidence intervals do not encompass the zero line.

Fig. \ref{fig:HPresults} reports the results of multiscale information storage analysis for the HP time series. Using the eAR method whereby long range correlations are not modeled (Fig. \ref{fig:HPresults}a), at scale 1 ($f_\tau=0.5$) the information stored in the HP process is significantly higher in the UP condition compared with SU. This reflects a widely known behavior of heart rate variability, whose complexity is known to decrease with head-up tilt due to an activation of the sympathetic nervous system which has a regularizing effect on the cardiac dynamics \cite{porta2017nonlinear,porta2007progressive}. Higher values of $S_X$ during orthostatic stress are observed also at increasing the time scale, and are detectable up to $f_\tau \sim 0.1$ Hz (Fig. \ref{fig:HPresults}b), reflecting the results obtained in Ref. \cite{FaesComplexity2017} where a lower multiscale entropy is detected in the same data. Here we see also that the tilt-induced increase of the information storage is present also when the slow trends affecting HP, likely due to long-range correlations, are filtered out using the eARd method (Fig.\ref{fig:HPresults}c); in this case the increase of $S_X$ from SU to UP is significant also for intermediate time scales (0.05 Hz $<f_\tau <$ 0.1 Hz, Fig. \ref{fig:HPresults}d)). This behavior is less evident when long range correlations are modeled, as $S_X$ increases during UP only at scale 1, and is even reverted at longer time scales, as $S_X$ decreases during UP for 0.01 Hz $<f_\tau <$ 0.1 Hz (Fig.\ref{fig:HPresults}e,f). This behavior, documenting that the complexity of heart rate variability \textit{increases} during tilt if observed at long time scales, has been previously observed using multiscale entropy 
\cite{turianikova2011effect}. Here, it becomes visible only modeling long-range correlations through the eARFI approach and indicates that long-range correlations are likely less important during head-up tilt. Thus, the utilization of the modeling approach proposed in this study suggests that postural stress augments the capability of HP to store information at low time scales but also diminish such capability at longer time scales.

Fig. \ref{fig:SAPresults} reports the results of multiscale information storage analysis for the SAP time series. According to the eAR method (Fig. \ref{fig:SAPresults}a,b), moving from SU to UP the index $S_X$ increases significantly at scale 1 ($f_\tau = 0.5$ Hz) and decreases significantly at scale 2 ($f_\tau = 0.25$ Hz), confirming in terms of information storage the results reported in Ref. \cite{FaesComplexity2017} based on a linear complexity measure. These two opposite behaviors of the information stored in the SAP process are here explained in terms of long-range correlations, which are removed or explicitly considered respectively through detrending or through performing ARFI identification. In fact, according to the eARd method (Fig. \ref{fig:SAPresults}c,d), $S_X$ is still significantly higher during UP when $f_\tau = 0.5$ Hz, but is not significantly different from SU for any other value of $f_\tau$. On the contrary, according to the eARFI method (Fig. \ref{fig:SAPresults}e,f), $S_X$ does not show significant differences between SU and UP when $f_\tau = 0.5$ Hz, but is significantly smaller when $f_\tau = 0.25$ Hz. These results suggest that the higher capability of SAP to store information during tilt observed at scale 1 is related exclusively to short-term dynamics, while the lower storage capability observed at intermediate scales ($f_\tau \sim 0.25$ Hz, where respiration-related components are suppressed) is driven by long-range correlations. Thus, head-up tilt induces scale-dependent variations in the complexity of arterial pressure, with higher complexity (lower $S_X$) associated with slow oscillations, and lower complexity (higher $S_X$) associated to the effects of respiration.

Fig. \ref{fig:RESPresults} reports the results of multiscale information storage analysis for the RESP time series. In this case we find that, using all methods, at short time scales the respiration process stores more information during UP than during SU, while at longer time scales the amount of information stored in the process does not change significantly with head-up tilt. 
This larger regularity of the respiration dynamics and its multiscale behavior confirm previous findings \cite{valente2018univariate, FaesComplexity2017}, further suggesting that long-range correlations do not have significant influence on the complexity of respiratory patterns.
These results may be expected since respiration is usually strongly evident in the so-called high-frequency band ($>0.15$ Hz) \cite{camm1996heart} and is thus filtered out almost entirely for time scales $\geq 2$.

\begin{figure}[ht]
\includegraphics[width=8.5 cm]{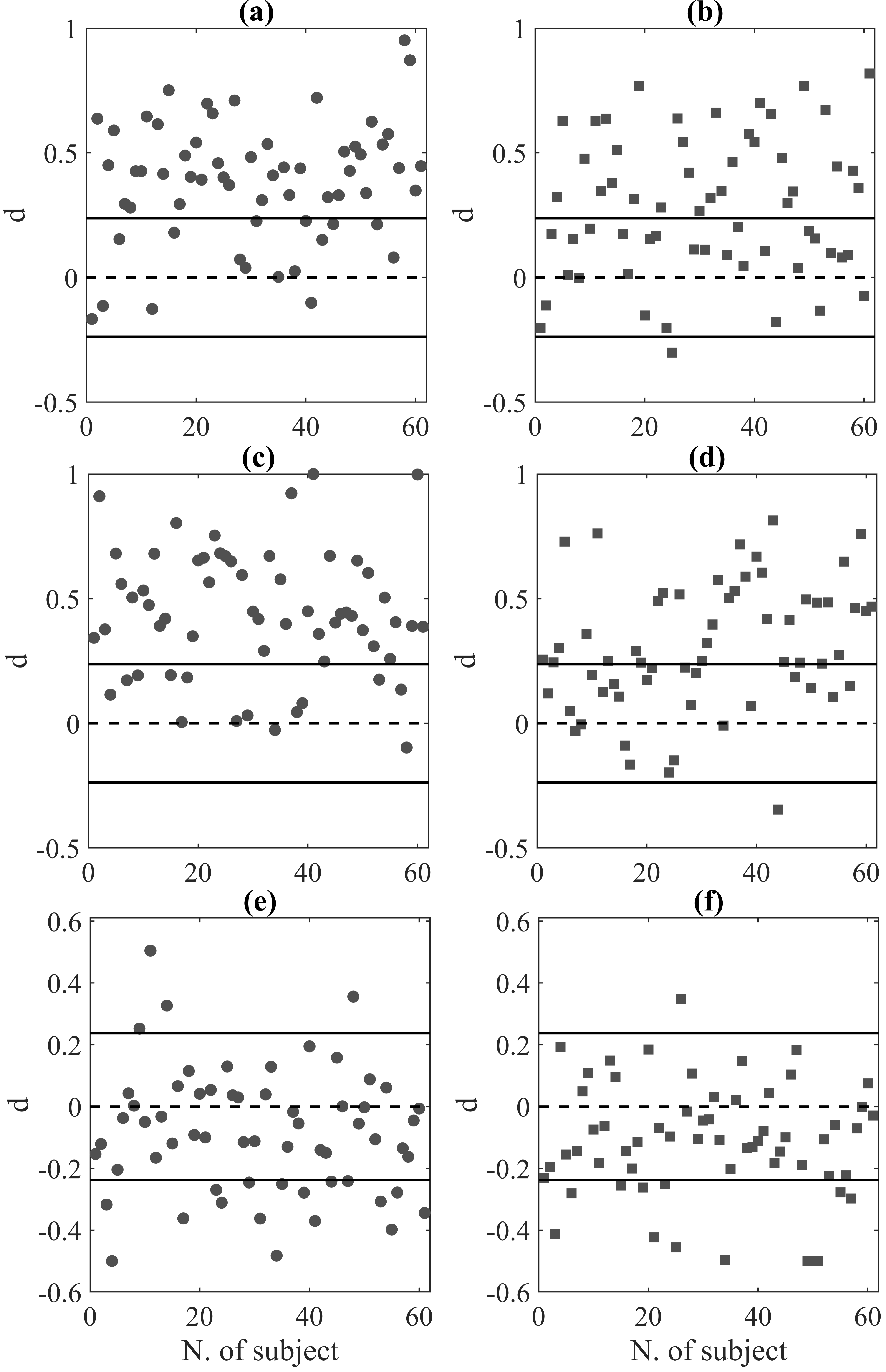}
\caption{\label{fig:dresults} Values of the differencing parameter $d$ estimated for each subject (index 1,...,61) for the heart period (a,b), systolic pressure (c,d) and respiration (e,f) time series in the supine (SU, a,c,e) and upright (UP, b,d,f) body positions. Each panel reports also the $95^{th}$ confidence intervals (solid lines) of the distribution of $d$ computed around the zero level (dashed line).}
\end{figure}

To further elucidate the role played by long-range correlations in determining the information stored in the considered processes, we analyze the values of the differencing parameter $d$ computed in the various conditions using the Whittle semi-parametric estimator \cite{Beran2012}. Fig. \ref{fig:dresults} reports, for any given process and condition, the individual values of $d$ plotted for each of the analyzed subjects, together with their $95^{th}$ confidence intervals relevant to the zero level (derived from the asymptotic statistic given by Eq. (6) of Ref. \cite{leite2013beyond}). We find that the differencing parameter computed for the HP series and for the SAP series is statistically significant (i.e., outside of the confidence intervals) in a lower number of subjects in the UP condition (Fig. \ref{fig:dresults}b,d) compared with SU (Fig. \ref{fig:dresults}a,c); this seems not to be the case for the RESP time series (Fig. \ref{fig:dresults}e vs. Fig. \ref{fig:dresults}f). These results are confirmed by the application of a Student t-test for paired data applied to the distributions of $S_X$ computed during SU and during UP, which returns $p-$values lower than the critical 0.05 level for HP and SAP (respectively $p=0.0037$ and $p=0.0192$), while the distributions were not significantly different for RESP ($p=0.377$).
These changes correspond to situations in which the eARFI method detects a statistically significant decrease of the information storage at intermediate/long time scales (Fig. \ref{fig:HPresults} and Fig. \ref{fig:SAPresults}). This suggests that when the importance of long-range correlations decreases for a time series (lower $d$), the slow dynamics of the series become more complex (lower $S_X$), supporting the argument that long-range correlations play a regularizing role for the process dynamics \cite{xiong2017entropy}.

\section{Conclusions}
We have introduced an approach to assess across multiple temporal scales the amount of information stored in a dynamic process, intended as the degree to which information is preserved in a time-evolving dynamical system in a way such that it can be retrieved from the past system states. Thanks to its parametric formulation, the proposed approach inherits the computational reliability of linear multiscale entropy \cite{FaesComplexity2017}, exploiting it for the assessment of regularity and – most importantly – allowing the simultaneous description of short-term dynamics and long memory properties. Our simulations show that the state space formulation implemented here, though being restricted to the description of linear dynamics, outperforms model-free multiscale complexity analysis \cite{Valencia20092202} and, thanks to the incorporation of long-range correlations, leads to a more reliable evaluation of the information storage at long time scales if compared with linear multiscale entropy \cite{FaesComplexity2017}.

Since long-range correlations are a fundamental aspect of multiscale dynamics, the present work opens the ways to a reliable computation of the dynamical complexity of several natural and man-made processes where different mechanisms coexist, operating across multiple temporal scales. Here, the application to cardiovascular dynamics led to unprecedented physiological results, such as the observation that, at temporal scales compatible with sympathetic neural activity, postural stress blunts the capability of heart rate and arterial pressure variability to actively store information.

Future developments should focus on extending the formulations proposed in this work to the multiscale representation of vector ARFI models, in order to attain a complete decomposition across time scales of the other constitutive elements of information processing in dynamical networks, i.e. information transfer and information modification \cite{Lizier2012,wibral2014local}.

\begin{acknowledgments}

Work partially supported by UID/MAT/00144/2013 (CMUP), UID/MAT/04106/2013 (CIDMA), which are funded by FCT with national (MEC) and European structural funds through the program FEDER, under PT2020, and project STRIDE-NORTE-01-0145-FEDER-000033 funded by ERDF-NORTE 2020.

\end{acknowledgments}



\bibliography{MSE_ARFI_biblio}

\end{document}